\documentclass[11pt]{article}

\usepackage[T1]{fontenc}
\usepackage[utf8]{inputenc}
\usepackage[english]{babel}
\usepackage{lmodern}
\usepackage{microtype}
\usepackage[letterpaper,margin=1in]{geometry}
\usepackage{amsmath,amssymb,amsthm,mathtools}
\usepackage{xurl}
\usepackage{booktabs}

\usepackage[hidelinks,breaklinks=true,hypertexnames=false]{hyperref}
\usepackage[nameinlink,noabbrev]{cleveref}

\hypersetup{
  pdftitle={Batch range mode in O(n+Q)\string^1.395},
  pdfauthor={Filippo di Gravina}
}

\theoremstyle{definition} 
\newtheorem{theorem}{Theorem}[section]
\newtheorem{lemma}[theorem]{Lemma}

\DeclareMathOperator{\modefreq}{modefreq}
\newcommand{\eps}{\varepsilon}

\allowdisplaybreaks
\title{Batch range mode in $O(n+Q)^{1.395}$}
\author{Filippo di Gravina\\
\small \textit{Affiliation:} Università degli Studi di Bari Aldo Moro\\
\small \textit{Email:} \texttt{f.digravina2@studenti.uniba.it}}
\date{October 8, 2026}

\begin{document}
\maketitle

\begin{abstract}
We give a deterministic worst-case algorithm for exact Batched Range Mode on an arbitrary array of length $n$ and a batch of $Q$ intervals. The algorithm returns the mode frequency and the smallest mode identifier for each query in time
\[
(n+Q)^{\tau+o(1)},\qquad
\tau=5-\sqrt{13}=1.394448724536\ldots.
\]
The algorithm combines virtual occurrence windows and block-boundary snapping with a query-restricted min-plus product whose first operand is column-monotone and whose second operand is row-monotone. A deterministic modular verifier uses geometric and query potentials to bound the same active regions; false modular matches are corrected by offline orthogonal range updates. The final analysis uses the sharper rectangular exponent curve of Uzna\'nski (2026), together with exact rational parameter balancing and a transfer to prime fields. The analysis covers every frequency scale and arbitrary batch sizes. Exact frequencies and the smallest mode identifiers are recovered without increasing the exponent.
\end{abstract}

\section{Introduction}\label{sec:introduction}

In Batched Range Mode, the input is an array and a list of intervals, and each interval asks for its most frequent color. We consider the exact offline problem, with no restrictions on the palette, the arrangement of occurrences, or the query intervals.

\begin{theorem}\label{thm:main}
Given an explicit array of length $n$ and an explicit batch of $Q$ intervals, the mode frequency and the smallest identifier among the modes of each nonempty interval can be computed in deterministic worst-case time
\[
(n+Q)^{\tau+o(1)},\qquad
\tau=5-\sqrt{13}=1.394448724536\ldots.
\]
\end{theorem}

Empty intervals have frequency zero and no mode color. Repeated intervals are allowed. Construction of intermediate instances, input reads, and output writes are included in the bound.

\paragraph{Previous work.}
Vassilevska Williams and Xu~\cite{WilliamsXu2020} connected Batched Range Mode to structured min-plus multiplication. Gu, Polak, Vassilevska Williams, and Xu~\cite{GuPolakVWXu2021} developed this connection using monotone min-plus products. Chi, Duan, Xie, and Zhang~\cite{ChiDuanXieZhang2022} improved the randomized monotone primitive, and D\"urr~\cite{Durr2023} extended it to rectangular dimensions; we use the corrected Batched Range Mode comparison in arXiv:2208.02862v4 rather than the erroneous numerical exponent in an earlier version. Jin, Park, Saha, and Xu~\cite{JinParkSahaXu2026} derandomized the monotone primitive and its downstream applications. Our product additionally uses monotonicity in both external coordinates and restricts correction to prescribed output cells. The matrix-multiplication analysis uses the rectangular exponent bound of Uzna\'nski~\cite{UznanskiRect2026}, based on the square-multiplication work of OpenAI~\cite{OpenAISquare2026}. For $Q=\Theta(n)$, our bound is $n^{5-\sqrt{13}+o(1)}$.

\paragraph{The product used here.}
Our first operand is monotone in its external coordinate, as is the second operand in the other external coordinate. Their min-plus product is therefore monotone in both output coordinates. We exploit this property while computing only a prescribed set $E$ of output cells. A recursive product on $E$ extends to a monotone candidate over the full grid using prefix maxima. This extension costs linear time in the grid size and does not require the true min-plus values elsewhere.

For the verifier, rounded blocks of the two operands define rectangles. Intersecting those rectangles with rounded candidate values gives regions that can be represented by three-dimensional boxes. The same active regions admit both a geometric bound and a bound charged to the requested cells. Combining these bounds across refinement levels is the source of the improved exponent. The modular selection follows the small-prime potential framework of Jin et al.~\cite{JinParkSahaXu2026}; we give the selection and correction arguments for the two potentials used here.

The Range Mode reduction uses virtual occurrence windows and a uniform grid of block boundaries. The colors in the two boundary fragments of a query are checked exactly. A witnessing color absent from those fragments has all its occurrences in the snapped interior. On that interior we use true prefix ranks, obtaining both operand monotonicities directly. This avoids a graph lifting or a completion of missing ranks.

\paragraph{Model and notation.}
We work on a word-RAM with $O(\log N)$-bit words, where $N=\max(2,n+Q)$. Color identifiers have $O(\log N)$ bits and are totally ordered. We write $\widetilde O$ for polylogarithmic factors and $\widehat O$ for $N^{o(1)}$ factors; all dimensions and integer magnitudes in intermediate instances are polynomially bounded in $N$. The uniform meaning of $N^{\tau+o(1)}$ is that for every $\eps>0$ there is an algorithm of cost $O_\eps(N^{\tau+\eps})$. We use exact ordinary matrix multiplication, fast polynomial evaluation and interpolation, and the prime-number theorem. The matrix-multiplication bounds and their finite-field implementation are justified in \cref{sec:matrix-bounds} using the explicitly cited external matrix-multiplication results. The finite-field and word-RAM implementation is proved there. Earlier rectangular bounds appear in~\cite{VWXXZ2024,ADVWXXZ2025}. The proof uses the cited rectangular matrix-multiplication bound as an external theorem and proves the transformations and asymptotic consequences required for the present algorithms. Rounding dimensions and cutoffs changes only constant or subpolynomial factors.

\section{A query-restricted two-sided monotone product}\label{sec:minplus}

Let $A\in\{0,\ldots,R\}^{a\times r}$ have nondecreasing columns, and let $B\in\{0,\ldots,R\}^{r\times b}$ have nondecreasing rows, where $a,b,r,R\ge1$. For a prescribed set $E\subseteq[a]\times[b]$ of $q\ge1$ cells, we seek
\begin{equation}\label{eq:minplus}
C^*_{ij}=\min_{k\in[r]}(A_{ik}+B_{kj}),\qquad (i,j)\in E.
\end{equation}
Let $M_0\ge ar+rb+ab$ be a valid deterministic cost for complete ordinary multiplication of these dimensions over sufficiently large prime fields. The set $E$ is supplied as an explicit list; duplicate cells can be coalesced by sorting.

\begin{theorem}[Two-sided monotone product]\label{thm:two-sided}
The values in \eqref{eq:minplus} can be computed in deterministic time
\begin{equation}\label{eq:two-sided-bound}
\widehat O\!\left(
M_0+q+(rR^2M_0^2)^{1/3}
+\sqrt{rR\min(R,\sqrt q)M_0}
\right).
\end{equation}
\end{theorem}

We prove the theorem through an equality verifier. Its candidate $C$ is a full $a\times b$ integer matrix, nondecreasing in both coordinates, with entries in $[0,4R]$. The verifier returns
\begin{equation}\label{eq:exact-verifier-count}
e_{ij}=\#\{k\in[r]:A_{ik}+B_{kj}=C_{ij}\},\qquad(i,j)\in E.
\end{equation}
The candidate need not agree with the true product outside $E$.

\subsection{Regions and their two potentials}\label{sec:regions}

Fix a dyadic value $\lambda=2^\ell$. For each $k$, partition the row indices into maximal intervals $I$ on which $\lfloor A_{ik}/\lambda\rfloor=x$ is constant, and partition the column indices into maximal intervals $J$ on which $\lfloor B_{kj}/\lambda\rfloor=y$ is constant. For each rectangle $I\times J$, with $I=[i_0,i_1]$ and $J=[j_0,j_1]$, introduce one region record
\[
\sigma=(k,I,J,x,y,c)
\]
for every integer
\[
\left\lfloor C_{i_0j_0}/\lambda\right\rfloor
\le c\le
\left\lfloor C_{i_1j_1}/\lambda\right\rfloor.
\]
The record represents the cells in $I\times J$ with $\lfloor C_{ij}/\lambda\rfloor=c$. Some records may represent no cells. Each cell belongs to exactly one record for each $k$. We retain empty records in the potentials so that the corner formulas below are exact.

\begin{lemma}\label{lem:region-count}
For $1\le\lambda\le R$, there are $O(rR^2/\lambda^2)$ region records.
\end{lemma}
\begin{proof}
Fix $k$. Since each quotient sequence is nondecreasing and the unrounded operands range over $[0,R]$, there are $s,t\le1+\lfloor R/\lambda\rfloor\le2R/\lambda$ quotient blocks. In each operand rectangle $I\times J$ with opposite corners $(i_0,j_0)$ and $(i_1,j_1)$, the number of labels is at most
\[
2+\frac{C_{i_1j_1}-C_{i_0j_0}}\lambda.
\]
Split the difference into
\[
(C_{i_1j_1}-C_{i_0j_1})+(C_{i_0j_1}-C_{i_0j_0}).
\]
For a fixed column block, the first differences over distinct row blocks are nonnegative increments on disjoint row intervals, so their sum is at most $4R$. Summing over the $t$ column blocks contributes at most $4Rt$. For a fixed row block, the second differences over the disjoint column blocks similarly sum to at most $4R$; this contributes at most $4Rs$. The total number of records for this $k$ is therefore at most
\[
2st+\frac{4R(s+t)}\lambda
\le 8(R/\lambda)^2+16(R/\lambda)^2
=24R^2/\lambda^2.
\]
Summing over $k$ proves the lemma. In particular the initial geometric and query budgets can be chosen as the explicit positive rational numbers
\[
 H^G_\ell=168rR^2/\lambda,\qquad
 H^E_\ell=7qr\lambda,
\]
since at most $6\lambda+1\le7\lambda$ integer shifts are charged per record or requested cell and color.
\end{proof}

Associate with each record the anchor
\[
\delta_\sigma=\lambda(x+y-c).
\]

For a fixed $k$, the monotone quotient sequences on both axes partition the grid into these rectangles. At any cell $(i,j)$ in such a rectangle, monotonicity of $C$ gives
\[
\bigl\lfloor C_{i_0j_0}/\lambda\bigr\rfloor
\le\bigl\lfloor C_{ij}/\lambda\bigr\rfloor
\le\bigl\lfloor C_{i_1j_1}/\lambda\bigr\rfloor.
\]
Consequently each triple $(i,j,k)$ belongs to exactly one region: its unique operand rectangle and the unique label $c=\lfloor C_{ij}/\lambda\rfloor$. This remains true even though the potential sums include empty regions. It is this uniqueness, rather than disjointness between different colors, that will be used in the charging argument.

Write the remainders of the three entries as
\[
 A_{ik}=\lambda x+\alpha_A,\quad
 B_{kj}=\lambda y+\alpha_B,\quad
 C_{ij}=\lambda c+\alpha_C,
 \qquad 0\le\alpha_A,\alpha_B,\alpha_C<\lambda.
\]
The exact gap $g=A_{ik}+B_{kj}-C_{ij}$ therefore satisfies
\begin{equation}\label{eq:exact-gap}
 g=\delta_\sigma+(\alpha_A+\alpha_B-\alpha_C),
 \qquad -\lambda<\alpha_A+\alpha_B-\alpha_C<2\lambda.
\end{equation}
If $m\mid g$ and $g\ne0$, choose $s=\delta_\sigma-g$. Then $|s|\le2\lambda$, $m\mid\delta_\sigma-s$, and $\delta_\sigma-s=g\ne0$. Thus every false modular equality is active at every dyadic level. When $\lambda=1$, the three remainders are zero and the anchor equals the exact gap.
For a modulus $m\ge1$, call the record active if
\begin{equation}\label{eq:active}
\exists s\in[-3\lambda,3\lambda]\cap\mathbb Z:
\quad m\mid\delta_\sigma-s,\qquad \delta_\sigma-s\ne0.
\end{equation}
Two divisions test this predicate by checking whether the interval
$[\lceil(\delta_\sigma-3\lambda)/m\rceil,
\lfloor(\delta_\sigma+3\lambda)/m\rfloor]$ contains a nonzero integer.

Define $Y^G_\ell(m)$ by counting all record--shift pairs satisfying divisibility in \eqref{eq:active}, including those with $\delta_\sigma=s$. Let $Z^G_\ell$ count the pairs with $\delta_\sigma=s$, and put $X^G_\ell(m)=Y^G_\ell(m)-Z^G_\ell$. For the query potential, use
\[
\delta_{ijk}^{\ell}
=\lambda\left(
\left\lfloor A_{ik}/\lambda\right\rfloor+
\left\lfloor B_{kj}/\lambda\right\rfloor-
\left\lfloor C_{ij}/\lambda\right\rfloor\right)
\]
and define $Y^E_\ell(m)$, $Z^E_\ell$, and $X^E_\ell(m)$ in the same way over $(i,j)\in E$, $k\in[r]$, and the shifts in $[-3\lambda,3\lambda]$. Both $Z$ terms are independent of $m$.

\begin{lemma}\label{lem:two-potentials}
There are bounds
\begin{equation}\label{eq:potential-budgets}
X^G_\ell(1)\le H^G_\ell=O(rR^2/\lambda),\qquad
X^E_\ell(1)\le H^E_\ell=O(qr\lambda).
\end{equation}
The number of active level-$\ell$ records containing a cell of $E$ is at most $\min(X^G_\ell(m),X^E_\ell(m))$.
\end{lemma}
\begin{proof}
There are at most $6\lambda+1$ shifts per record or query--color pair, so the explicit record bound from \cref{lem:region-count} proves the first inequality, while direct counting proves the second. An active record has at least one nonzero shifted witness counted by $X^G$. For each active region meeting $E$, choose one requested cell and one shift witnessing activity. The anchor at that cell and color agrees with the region anchor. Distinct records cannot be mapped to the same triple $(i,j,k,s)$: by the uniqueness established above, a fixed $(i,j,k)$ belongs to only one record at this level. Hence these witnesses inject into the query potential and the number of active nonempty regions is at most $X^E$. The geometric potential independently bounds the same regions.
\end{proof}

\subsection{Computing both potentials}\label{sec:potential-products}

For this subsection fix $\ell,m$, and write
\[
u_{ik}=\lambda\lfloor A_{ik}/\lambda\rfloor,\qquad
v_{kj}=\lambda\lfloor B_{kj}/\lambda\rfloor.
\]
In the ring modulo $z^m-1$, set $\mathcal A_{ik}=z^{u_{ik}\bmod m}$ and $\mathcal B_{kj}=z^{v_{kj}\bmod m}$. Let $\mathcal A^{\mathrm{beg}}$ and $\mathcal A^{\mathrm{end}}$ retain these monomials only at the first and last indices of each column block of $A$; define $\mathcal B^{\mathrm{beg}}$ and $\mathcal B^{\mathrm{end}}$ at the endpoints of the row blocks of $B$. Compute
\[
D=\mathcal A\mathcal B,\qquad
D^{\mathrm{beg}}=\mathcal A^{\mathrm{beg}}\mathcal B^{\mathrm{beg}},\qquad
D^{\mathrm{end}}=\mathcal A^{\mathrm{end}}\mathcal B^{\mathrm{end}}.
\]
Write $D_{ijt}$ for the coefficient of $z^t$, and similarly for the other products. Put
\[
w_{\lambda,m}(e)=\#\{s\in[-3\lambda,3\lambda]\cap\mathbb Z:s\equiv e\pmod m\},\qquad
W_{\lambda,m}(t,c)=\sum_{h=0}^{c}w_{\lambda,m}(t-\lambda h),
\]
where $W(t,-1)=0$. With $c_{ij}=\lfloor C_{ij}/\lambda\rfloor$, the query potential is
\begin{equation}\label{eq:query-potential-product}
Y^E_\ell(m)=\sum_{(i,j)\in E}\sum_{t=0}^{m-1}
D_{ijt}\,w_{\lambda,m}(t-\lambda c_{ij}).
\end{equation}
For the geometric potential, every operand rectangle contributes the cumulative weight at its upper-right corner minus the cumulative weight immediately before its lower-left corner:
\begin{equation}\label{eq:corner-potential}
Y^G_\ell(m)=\sum_{i,j}\sum_{t=0}^{m-1}
\left(
D^{\mathrm{end}}_{ijt}W_{\lambda,m}(t,c_{ij})-
D^{\mathrm{beg}}_{ijt}W_{\lambda,m}(t,c_{ij}-1)
\right).
\end{equation}
To verify the identity, fix $k$ and one operand rectangle with rounded sum $u+v$, and put $c_-=\lfloor C_{i_0j_0}/\lambda\rfloor$, $c_+=\lfloor C_{i_1j_1}/\lambda\rfloor$. Its contribution is exactly
\[
\sum_{c=c_-}^{c_+}w_{\lambda,m}(u+v-\lambda c)
=W_{\lambda,m}((u+v)\bmod m,c_+)
-W_{\lambda,m}((u+v)\bmod m,c_--1).
\]
The endpoint-masked products count this rectangle's monomial precisely once at $(i_1,j_1)$ and once at $(i_0,j_0)$, respectively. Summing over all $k$ and their operand rectangles gives \eqref{eq:corner-potential}. The formula counts all labels in the corner interval, including labels whose regions are empty. Likewise $D_{ijt}$ counts exactly the indices $k$ whose rounded sum has residue $t$, proving \eqref{eq:query-potential-product}.

The cumulative weight can be evaluated without a table of size $m^2$. For $c\ge0$,
\begin{equation}\label{eq:floor-kernel}
W_{\lambda,m}(t,c)=\sum_{h=0}^{c}
\left(
\left\lfloor\frac{t-\lambda h+3\lambda}{m}\right\rfloor-
\left\lfloor\frac{t-\lambda h-3\lambda-1}{m}\right\rfloor
\right).
\end{equation}
Each floor sum is computable in $O(\log N)$ arithmetic operations by the following exact Euclidean routine. Define
\[
S(n,m,a,b)=\sum_{j=0}^{n-1}\left\lfloor\frac{aj+b}{m}\right\rfloor,
\quad n\ge0,\ m\ge1,
\]
for signed integer $a,b$. Euclidean division $a=q_am+a_0$ and $b=q_bm+b_0$, with $0\le a_0,b_0<m$, yields the identity
\begin{equation}\label{eq:floor-reduce}
S(n,m,a,b)=q_a\frac{n(n-1)}2+q_bn+S(n,m,a_0,b_0).
\end{equation}
After adding these quotient contributions, if $a_0n+b_0<m$ the remaining sum is zero. Otherwise let $y=a_0n+b_0$ and replace
\[
(n,m,a_0,b_0)\longleftarrow
(\lfloor y/m\rfloor,a_0,m,y\bmod m).
\]
Then repeat the quotient extraction. The replacement is valid by counting integer lattice points under $a_0j+b_0$: for $0\le a_0,b_0<m$ and $a_0n+b_0\ge m$,
\begin{equation}\label{eq:floor-swap}
S(n,m,a_0,b_0)=S(\lfloor y/m\rfloor,a_0,m,y\bmod m).
\end{equation}
To derive \eqref{eq:floor-swap}, count for each $t\ge1$ the indices $0\le j<n$ satisfying $a_0j+b_0\ge mt$, reverse the summations, and use $y=a_0n+b_0$ for the last partial strip. In this case $a_0>0$, so the new denominator is nonzero. After the next quotient extraction the modulus decreases as in Euclid's algorithm. This proves logarithmically many steps and supports negative coefficients through \eqref{eq:floor-reduce}. Using $(a,b)=(-\lambda,t+3\lambda)$ and $(-\lambda,t-3\lambda-1)$ evaluates \eqref{eq:floor-kernel} exactly.
The three polynomial products and the sums in \eqref{eq:query-potential-product} and \eqref{eq:corner-potential} cost $\widehat O(mM_0)$. For exact implementation, evaluate the ordinary polynomial product, of degree below $2m$, at $O(m)$ field points, apply ordinary scalar matrix multiplication, interpolate, and reduce modulo $z^m-1$. Its cyclic coefficients are counts at most $r$. A prime characteristic greater than $\max(D_\eps,r,2m)$ therefore suffices; a sieve finds a suitable prime of magnitude $O(D_\eps+r+m)$, where $D_\eps$ is the accuracy-dependent constant from \cref{lem:field-transfer}. For fixed accuracy its search cost is covered by the stated bound. The coefficient counts are recovered as integers before evaluating the weighted sums. Evaluation, interpolation, and the coefficient scans cost $\widetilde O(m(ar+rb+ab))$, covered by $\widehat O(mM_0)$. All subsequent integer magnitudes have $O(\log N)$ bits.

\subsection{Deterministic selection and refinement}\label{sec:modulus}

Choose $1\le P\le R$ and consider $0\le\ell\le\lfloor\log_2P\rfloor$. We select one modulus simultaneously for the two potentials at every level. Let $\nu$ index these $O(\log N)$ potentials, with initial bounds $H_\nu$ from \eqref{eq:potential-budgets}. Choose
\[
K=\lceil\log_2N\rceil,\qquad
L=2^{\left\lceil\sqrt{K/\max(1,\lceil\log_2K\rceil)}\right\rceil}
\]
and sieve the family $\mathcal P$ of primes in $[L/2,L]$. Its size is $\Theta(L/\log L)$; bounded small instances are handled directly.

Start with $m_0=1$ and suppose $X_\nu(m_t)\le H_{\nu,t}$ for every potential. Each surviving witness is a nonzero integer $z$ divisible by $m_t$, with $|z|\le N^{C_0}$ for an absolute $C_0$ fixed by the polynomial bounds on instance magnitudes. For every trial prime $p$, including one previously chosen, we have
\[
 m_tp\mid z\quad\Longleftrightarrow\quad p\mid z/m_t.
\]
If $s$ distinct primes from $[L/2,L]$ divide the nonzero integer $z/m_t$, then
\[
 (L/2)^s\le |z/m_t|\le N^{C_0},
 \qquad s\le\frac{C_0\log N}{\log(L/2)}.
\]
The prime-number theorem gives $|\mathcal P|\ge c_0L/\log L$ for all sufficiently large $N$. The fraction of trial primes dividing any fixed quotient $z/m_t$ is thus at most $c_1K/L$ for an absolute constant $c_1$. Double counting the surviving witnesses proves
\[
 \frac1{|\mathcal P|}\sum_{p\in\mathcal P}
 \frac{X_\nu(m_tp)}{H_{\nu,t}}\le c_1K/L
 \qquad\text{for every }\nu.
\]
There are $J=2(1+\lfloor\log_2P\rfloor)=O(K)$ potentials, so summing the normalized inequalities and choosing one prime no worse than the mean shows that a single $p_*$ satisfies all $J$ contractions
\[
 X_\nu(m_tp_*)\le\eta H_{\nu,t}\quad(\text{all }\nu),
 \qquad\eta=cK^2/L,
\]
where $c$ is a sufficiently large fixed integer. This is a deterministic existence argument and uses no independence assumption. Integer comparisons of squares suffice to compute $L$; the quantity $\eta$ is rational, and $2\eta<1$ for sufficiently large $N$. Bounded smaller $N$ are handled directly.
This prime can be found without knowing $Z_\nu$. Compute all $Y_\nu(m_tp)$, let $Y_\nu^{\min}=\min_pY_\nu(m_tp)$, and minimize
\[
\Phi(p)=\max_\nu
\frac{Y_\nu(m_tp)-Y_\nu^{\min}}{H_{\nu,t}}.
\]
The prime $p_*$ has $\Phi(p_*)\le\eta$ because the differences cancel $Z_\nu$ and $X_\nu\ge0$. For the selected prime $p_t$,
\[
\frac{X_\nu(m_tp_t)}{H_{\nu,t}}
\le\frac{\min_pX_\nu(m_tp)}{H_{\nu,t}}+\Phi(p_t)
\le2\eta.
\]
Set $m_{t+1}=m_tp_t$ and $H_{\nu,t+1}=2\eta H_{\nu,t}$. Stop when $m_T\ge P$. Then $P\le m_T<PL$, $m_T\le L^T$, and $T=O(1+\log P/\log L)$. Indeed,
\[
 m_T(2\eta)^T\le(2\eta L)^T
 \le\exp\left(O\left(\frac{K\log K}{\log L}+\log K\right)\right)
 =N^{o(1)},
\]
because $\log L=\Theta(\sqrt{K/\log K})$. Consequently
\begin{equation}\label{eq:contracted-potential}
X_\nu(m_T)\le H_\nu(2\eta)^T
\le\frac{H_\nu}{m_T}N^{o(1)}.
\end{equation}
Each stage costs $\widehat O(L^2m_tM_0)$ by the potential formulas. The pretermination values satisfy $m_t<P$, so selection costs $\widehat O(PM_0)$. All normalized comparisons are exact rational comparisons. Indeed $H_{\nu,t}=H_\nu(2cK^2/L)^t$ and $T=O(1+K/\log L)$, so numerators and denominators have $O(K+T\log K+T\log L)=O(K)$ bits. The explicit positive rational budgets from \cref{lem:region-count} satisfy every initial inequality and make the deterministic objective well-defined.

\begin{lemma}\label{lem:refinement}
An active region at level $\ell$ has an active parent at level $\ell+1$. Each parent has at most eight children.
\end{lemma}
\begin{proof}
Write the child's block size as $\lambda$ and the parent's as $2\lambda$. Each parent operand block splits into at most two child blocks, and a parent candidate label splits into the two labels $2c,2c+1$. Thus there are at most eight children. The child labels are $2x+e_A$, $2y+e_B$, and $2c+e_C$, where each $e$ is zero or one. Hence
\[
\delta_{\mathrm{child}}-\delta_{\mathrm{parent}}
=\lambda(e_A+e_B-e_C)\in[-\lambda,2\lambda].
\]
If $\delta_{\mathrm{child}}-s=z\ne0$ is divisible by $m_T$ with $|s|\le3\lambda$, the parent has the same witness $z$ with a shift of magnitude at most $5\lambda\le3(2\lambda)$. A child containing a requested cell has a parent containing that cell.
\end{proof}

Construct a deterministic orthogonal range tree on the $q$ points $(i,j,C_{ij})$, $(i,j)\in E$. Its construction costs $\widetilde O(q)$, and emptiness of a three-dimensional box is tested in polylogarithmic time. A region record represents the box
\[
I\times J\times[c\lambda,(c+1)\lambda-1].
\]
At the top level $\lambda=2^{\lfloor\log_2P\rfloor}=\Theta(P)$, enumerate all records and retain those that are active and nonempty on the query points. By \cref{lem:region-count}, this costs $\widetilde O(ar+rb+ab+q+rR^2/P^2)$. Refine retained records only, testing their at most eight children for activity and query emptiness. Operand block splits are found by binary search; candidate corner labels have constant-time access.

By \cref{lem:two-potentials} and \eqref{eq:contracted-potential}, the number of retained records at level $\lambda$ is
\begin{equation}\label{eq:joint-active-bound}
\widehat O\!\left(
\min\left\{\frac{rR^2}{P\lambda},\frac{qr\lambda}{P}\right\}
\right).
\end{equation}
Summing over dyadic levels gives
\begin{equation}\label{eq:refinement-cost}
\widehat O\!\left(\frac{rR\min(R,\sqrt q)}P\right).
\end{equation}
For $q\le R^2$, split the sum at $\lambda=R/\sqrt q$; the smaller-scale query bounds and the larger-scale geometric bounds are geometric series of total $O(rR\sqrt q/P)$. If the split lies above the top level, the query bounds alone are smaller still. For $q>R^2$, the geometric bounds sum to $O(rR^2/P)$.

\subsection{Exact correction}\label{sec:false-matches}

One further polynomial product, using the unrounded exponents of $A$ and $B$, gives
\[
b_{ij}=\#\{k:m_T\mid A_{ik}+B_{kj}-C_{ij}\},\qquad(i,j)\in E,
\]
in $\widehat O(PM_0)$ time. At level zero, the operands and the candidate are constant on a region, and its anchor is the exact integer gap. Every false modular match belongs to an active record containing its query point, using shift zero. By \cref{lem:refinement}, that record is recovered. For each recovered record whose gap is a nonzero multiple of $m_T$, add one to the correction count on its three-dimensional box. Records with zero gap are never corrected.

These box updates do not require enumerating the query points they contain. For an integer interval, use the identity
\[
\mathbf 1_{L\le x\le U}
=\mathbf 1_{x\ge L}-\mathbf 1_{x\ge U+1}.
\]
Multiplying these identities for the three coordinates of $I\times J\times[c,c]$ expands one update into eight signed dominance corners $(a,b,c')$. A corner adds its sign to each query point $(i,j,C_{ij})$ with $i\ge a$, $j\ge b$, and $C_{ij}\ge c'$. Sort corners and query points in increasing first coordinate, processing corners before queries on a tie. Maintain point additions and prefix-orthant sums on the two remaining coordinates by a deterministic coordinate-compressed, nested Fenwick tree. The offline processing and storage cost are $O((U+q)\log^2(U+q))$ for $U$ boxes. Thus every query obtains exactly the number $f_{ij}$ of false matches. A fixed color contributes once: $(i,j,k)$ belongs to exactly one level-zero record, and its gap is zero or a nonzero multiple of $m_T$ precisely in the two respective cases. Subtracting the second count from the modular count gives $e_{ij}=b_{ij}-f_{ij}$, which proves \eqref{eq:exact-verifier-count}.

Combining selection, initialization, refinement, and correction proves the verifier bound
\begin{equation}\label{eq:verifier-cost}
\widehat O\!\left(
M_0+q+PM_0+\frac{rR^2}{P^2}
+\frac{rR\min(R,\sqrt q)}P
\right).
\end{equation}
The $M_0$ term covers input reads and the full candidate grid.

\subsection{Recovering the product on the requested cells}\label{sec:verification-to-product}

Round both operands down after division by two and recursively compute their min-plus product $C'$ only on $E$. The true product of the rounded operands is nondecreasing in both coordinates. Its requested values therefore extend consistently to
\begin{equation}\label{eq:candidate-completion}
\overline C'_{ij}=\max\left(
\{0\}\cup\{C'_{xy}:(x,y)\in E, x\le i, y\le j\}
\right).
\end{equation}
At a requested cell, every term in this maximum is at most that cell's recursive value, and the value at the cell itself is included. Thus the extension agrees with $C'$ on $E$. A two-dimensional prefix-maximum scan constructs it in $O(ab+q)$ time.

For each $k$ write $A_{ik}=2\lfloor A_{ik}/2\rfloor+e_A$ and $B_{kj}=2\lfloor B_{kj}/2\rfloor+e_B$, where $e_A,e_B\in\{0,1\}$. Thus every sum lies between twice its rounded sum and twice that sum plus two. Taking minima over $k$ proves that at each $(i,j)\in E$ the original minimum lies in $\{2C'_{ij},2C'_{ij}+1,2C'_{ij}+2\}$. Verify the three full-grid candidates $2\overline C'+s$, $s=0,1,2$, and take the smallest with a positive equality count at that requested cell. An accepted value is an actual sum, and one of the candidates is the minimum. The candidates are nondecreasing in both coordinates and lie in $[0,4R]$. The base case is zero once both operands are zero. There is one recursive child per level, the requested set stays fixed, and the depth is $O(\log(R+1))$.

Put $L_0=rR^2$ and $K_0=rR\min(R,\sqrt q)$. In \eqref{eq:verifier-cost}, choose $P$ within a constant factor of
\[
\max\{1,(L_0/M_0)^{1/3},(K_0/M_0)^{1/2}\}.
\]
The unrounded value is at most $R$ because $M_0\ge r$ and $K_0\le rR^2$; rounding and capping at $R$ change costs by only a constant factor. The three parameter-dependent terms are bounded by
\[
O\!\left(M_0+(L_0M_0^2)^{1/3}+\sqrt{K_0M_0}\right).
\]
At lower recursion levels, the dimensions and $E$ are unchanged and the range decreases. The logarithmic recursion overhead is absorbed in $\widehat O$, proving \cref{thm:two-sided}.

\section{Threshold queries by virtual windows and block snapping}\label{sec:thresholds}

Define the known-threshold problem as follows. For each query interval $[l_t,r_t]$ and positive integer threshold $k_t$, decide whether some color occurs at least $k_t$ times in the interval. A positive answer need not yet identify the color. Group the queries by the dyadic scale
\[
v=2^{\lfloor\log_2 k_t\rfloor},\qquad v\le k_t<2v.
\]
A threshold larger than the interval length is immediately false. There are $O(\log n)$ scales and the total number of query records over all scales is $Q$.

\subsection{Virtual occurrence windows}\label{sec:virtual-windows}

Fix one scale $v$ and consider queries with
\begin{equation}
v\le k_q<2v.
\label{eq:L6}
\end{equation}
For an original color with occurrence positions $p_1<\cdots<p_m$, $m\ge v$, create virtual windows beginning at occurrence indices $1,v+1,2v+1,\ldots$, as long as at least $v$ occurrences remain. A window beginning at local index $s$ consists of occurrences $p_s,\ldots,p_{\min(m,s+3v-1)}$ and is treated as a new virtual color carrying the identifier of its original color.

There are at most $n/v$ virtual windows, and each original occurrence belongs to at most three windows. Hence the total number of virtual occurrences is at most $3n$.

Every run of $k$ consecutive occurrences with $v\le k<2v$ is contained in one of these windows. If the first occurrence in the run has local index $i$, choose the window beginning at
\[
1+v\left\lfloor\frac{i-1}{v}\right\rfloor.
\]
The run ends before the beginning index plus $3v$, and the chosen beginning is valid because at least $k\ge v$ occurrences remain. Therefore, for every query interval,
\begin{equation}
\max_d f_d[l,r]\ge k
\quad\Longleftrightarrow\quad
\exists\text{ virtual color }c:\ f_c[l,r]\ge k,
\qquad v\le k<2v.
\label{eq:L7}
\end{equation}
The reverse implication is immediate because a window contains only occurrences of one original color. The forward implication uses a window containing $k$ consecutive occurrences of the witnessing original color. No regularity of the occurrence distribution is required.

\subsection{Exact boundary processing and snapped interiors}\label{sec:snapping}

Choose an integer block length $D\ge1$ and let
\[
G=\{0,D,2D,\ldots,\lfloor n/D\rfloor D\}\cup\{n\},
\]
with duplicates removed. Consecutive grid boundaries have distance at most $D$. For a query $I=[l,r]$, put $L=l-1$ and $R_q=r$ and define
\[
u=\min\{g\in G:g\ge L\},\qquad
w=\max\{g\in G:g\le R_q\}.
\]
If $u<w$, the interior is $J=[u+1,w]$; otherwise let $J$ be empty. In all cases $I\setminus J$ consists of at most two fragments and contains fewer than $2D$ positions. When $u>w$, both query boundaries lie in a single grid gap, so the whole query has length less than $D$; when $u=w$, the two fragments together have length less than $2D$.

Build the occurrence list of every original color. For every query, inspect all positions of $I\setminus J$. For each color encountered, count its occurrences in the whole query $I$ by two binary searches in its occurrence list. Let $b_I$ be the largest count obtained, or zero if the fragments are empty. Repeated colors may be inspected repeatedly: the bound remains $\widetilde O(n+QD)$, including list construction. This processing is performed once for the entire threshold batch.

\begin{lemma}[Snapping identity]\label{lem:snapping}
For every positive threshold $k$,
\[
\modefreq(I)\ge k
\quad\Longleftrightarrow\quad
b_I\ge k\ \text{ or }\ \modefreq(J)\ge k.
\]
\end{lemma}
\begin{proof}
Both right-hand witnesses are actual colors in $I$. Conversely, a witnessing color either occurs in $I\setminus J$, in which case its exact count is among the boundary counts, or has all its occurrences in $J$. This includes empty interiors.
\end{proof}

For a virtual color $c$, let $F_c(t)$ be its true prefix count at coordinate $t$, with $0\le F_c(t)\le3v$. Use all grid boundaries as rows and as columns. Set
\[
A_{gc}=F_c(g),\qquad B_{cg}=3v-F_c(g),
\]
and reverse the column order of $B$. Then $A$ has nondecreasing columns and $B$ has nondecreasing rows. For the cell whose row is $u$ and whose reversed column represents $w$,
\begin{equation}\label{eq:snapped-frequency}
3v-\min_c(A_{uc}+B_{cw})=\max_c\bigl(F_c(w)-F_c(u)\bigr)
=\max_c f_c(J).
\end{equation}
Use \cref{thm:two-sided} only on the distinct cells requested by nonempty interiors at this scale. Store the original query identifiers and their thresholds in lists attached to the cells. Duplicate coalescing costs $\widetilde O(Q_v)$, and testing all thresholds afterwards costs $O(Q_v)$. If there are no virtual colors, every interior threshold at this scale is false. Otherwise \eqref{eq:L7}, \eqref{eq:snapped-frequency}, and \cref{lem:snapping} prove exactness.

The grid has $a=b\le n/D+2$ boundaries, the number of virtual colors is $r\le n/v$, and $q\le Q_v$. Constructing the occurrence windows takes $O(n)$ time. Scanning their sorted lists against the grid builds the dense rank operands in $O(n+ar+rb)$ time. A full candidate grid costs $O(ab)$ and is charged to $M_0$. No regularity of query endpoints or of color occurrence positions is assumed.

\section{Rectangular multiplication bounds and finite-field transfer}\label{sec:matrix-bounds}

For a real $0\le\beta\le1$, write $\omega(1,\beta,1)$ for the arithmetic exponent of multiplying an $x\times\lceil x^\beta\rceil$ matrix by a $\lceil x^\beta\rceil\times x$ matrix over the complex numbers. We use the stated rectangular exponent theorem and establish the finite-field algorithms required for the word-RAM complexity analysis.

\subsection{The external rectangular multiplication theorem}\label{sec:rect-external}

Our asymptotic result uses the following \emph{external theorem}.  Theorem~1.1 of Uzna\'nski~\cite{UznanskiRect2026} establishes, for arithmetic matrix multiplication over $\mathbb C$, the upper bound
\begin{equation}\label{eq:improved-envelope}
f(\beta)=\begin{cases}
2,&0\le\beta\le1/2,\\
1+\beta+1/(4\beta),&1/2\le\beta\le1.
\end{cases}
\end{equation}
This is an unconditional upper bound on $\omega(1,\beta,1)$, conditional only on the correctness of the cited external proof, and not a conjecture on the true exponent. Its continuous first derivative on $[0,1]$ is $0$ for $\beta\le1/2$ and $1-1/(4\beta^2)$ for $\beta\ge1/2$; the second derivative on the latter interval is $1/(2\beta^3)>0$.  In particular, $f$ is continuous, nondecreasing, and convex.  Only this rectangular exponent bound is used below.

The arXiv version v4 of D\"urr~\cite{Durr2023} corrects an earlier numerical comparison for Batched Range Mode.  In order not to conflate historical versions, the bibliographical entry for Jin, Park, Saha, and Xu~\cite{JinParkSahaXu2026} identifies exactly the arXiv version used here. Neither previous numerical bound enters the proof of the present theorem.

\subsection{Exact transfer and uniformity over all shapes}

\begin{lemma}[Prime-field implementation]\label{lem:field-transfer}
For every $\eps>0$ there is an integer $D_\eps\ge1$ such that, for every prime $p>D_\eps$ and every $0\le\beta\le1$, multiplication of an $x\times\lceil x^\beta\rceil$ matrix by a $\lceil x^\beta\rceil\times x$ matrix over $\mathbb F_p$ has a deterministic algorithm costing
\[
O_\eps(x^{f(\beta)+\eps})
\]
base-field operations. The algorithm and bound are uniform over $x,\beta,p$. For polynomially bounded $p,x$, word-RAM implementation introduces only polylogarithmic overhead.
\end{lemma}
\begin{proof}
Fix an accuracy smaller than $\eps$. Select a finite mesh of aspect ratios on $[0,1]$, fine enough that padding to the next larger mesh ratio raises $f$ by less than $\eps/4$; include $1/2$ and $1$. At every mesh point the definition of the matrix-multiplication exponent supplies a fixed finite exact bilinear rank scheme whose recursive blocking and padding achieve its exponent with slack less than $\eps/4$. This is the standard finite-scheme characterization of the exponent (Lemma~2.1 of \cite{OpenAIRect2026}); for the square result it is also the recursive algorithm at the end of Section~5 of \cite{OpenAISquare2026}. Dimensions and coefficients of these finitely many schemes depend only on the accuracy.

Their coefficient identities form finite polynomial systems over $\mathbb Q$. A complex solution implies a solution over $\overline{\mathbb Q}$, by the Nullstellensatz, so all schemes can be realized over a single number field $K$. Choose a $\mathbb Q$-basis $e_1=1,e_2,\ldots,e_c$ of $K$ and an integer $D_\eps$ clearing the denominators of all scheme coefficients and of the basis multiplication table. Increasing $D_\eps$ if necessary, their coefficients lie in the unital algebra
\[
\mathcal A=\bigoplus_{i=1}^c\mathbb Z[1/D_\eps]e_i,
\]
which is closed under multiplication. For $p>D_\eps$, reduce this free algebra modulo $p$. It is a $c$-dimensional commutative algebra over $\mathbb F_p$, and every exact scheme identity survives reduction. It need not be a field.

Run each whole recursive bilinear circuit in this algebra on inputs in the component of $e_1$. Every addition or multiplication is simulated by a fixed number of base-field operations, using the fixed multiplication table. Correctness of the reduced identities puts the desired output in the component of $e_1$; projecting onto that component recovers the exact product. This is a constant factor on the entire circuit, rather than a factor in its rank at every recursion depth. The finite mesh makes the algebra dimension and excluded denominators independent of the input shape and size.

Padding to the next mesh point and choosing the accuracy slack sufficiently small yields the displayed bound. The finitely many constant-size schemes and algebra tables are part of the fixed algorithm for that accuracy. Modular operations on $O(\log N)$-bit quantities take polylogarithmic word-RAM time. The verifier can choose $p>\max(D_\eps,r,2m)$ of polynomial magnitude by a sieve; all bounded small cases are absorbed by the accuracy-dependent constant. No approximate complex arithmetic is used.
\end{proof}

\section{The all-scales running time}\label{sec:global-cost}

Set
\begin{equation}\label{eq:parameters}
\begin{aligned}
d&=\frac{394449}{1000000},\qquad h=1-d=\frac{605551}{1000000},\\
\tau_0&=1+d=\frac{1394449}{1000000}.
\end{aligned}
\end{equation}
Thus $d=0.394449$, $h=0.605551$, and $\tau_0=1.394449$. This fixed rational choice provides an explicit strict certificate; the limit of the same estimates at nearby rational choices will yield $\tau=5-\sqrt{13}$. Choose the integer block length $D=\lceil N^d\rceil$. This choice is computable exactly: if $d=u/z$, binary search for the least integer $D$ with $D^z\ge N^u$. The fixed exponents $u,z$ increase constants only, and integer arithmetic has polylogarithmic cost. Write $v=N^\alpha$ for a scale. Constant factors and polylogarithmic overheads contribute $o(1)$ to the exponents. The boundary processing costs $\widetilde O(n+QD)\le N^{1+d+o(1)}$ once for the threshold batch.

\subsection{The outer scales}

For $\alpha\le d$, enumerate every consecutive subwindow of every virtual color's occurrence list. Each list has length at most $3v$ and their total length is at most $3n$, so there are $O(nv)$ subwindows. A subwindow from position $p_i$ to $p_j$ has weight $j-i+1$ and lies in a query $[l,r]$ exactly when $l\le p_i$ and $p_j\le r$. Sort subwindows and queries by their first endpoint, sweep, and maintain prefix maxima of weights indexed by the second endpoint. This computes the maximum contained subwindow weight in $\widetilde O(nv+Q_v)$ time. The occurrences of a color inside a query form a consecutive subwindow, so the returned maximum is its exact virtual mode frequency and \eqref{eq:L7} supplies the threshold answers. The cost is at most $N^{1+d+o(1)}$. Use the original intervals at these scales; snapping is unnecessary.

For $\alpha\ge h$, scan the at most $n/v$ virtual colors for each scale query. Binary searches give each count, in time $\widetilde O(n+Q_vn/v)\le N^{2-\alpha+o(1)}\le N^{1+d+o(1)}$. Again use the original intervals.

\subsection{The intermediate scales and exact balancing}

For $d\le\alpha\le h$, use the snapped interiors and true rank operands of \cref{sec:snapping}. Their dimensions satisfy
\[
a,b\le O(N^h),\qquad r\le N^{1-\alpha},\qquad
q\le Q_v\le N,\qquad R=3v=O(N^\alpha).
\]
The input and full-grid volumes are at most $O(N^{2h})$. Padding the outer dimensions and applying \cref{lem:field-transfer} gives a valid ordinary multiplication cost
\begin{equation}\label{eq:matrix-exponent}
M_0\le N^{M(\alpha)+o(1)},\qquad
M(\alpha)=h f\!\left(\frac{1-\alpha}{h}\right)
=h+1-\alpha+\frac{h^2}{4(1-\alpha)}.
\end{equation}
The aspect ratio belongs to $[d/h,1]\subset(1/2,1]$, so the second branch of \eqref{eq:improved-envelope} gives the displayed rational formula. Also $f\ge2$ covers $ar+rb+ab$. In \cref{thm:two-sided}, the remaining two terms have exponents at most
\begin{equation}\label{eq:heavy-exponents}
T_3(\alpha)=\frac{1+\alpha+2M(\alpha)}3,\qquad
T_2(\alpha)=\frac{1+\min(\alpha,1/2)+M(\alpha)}2.
\end{equation}

The function $M$ is convex and nonincreasing. Hence its maximum is $M(d)$, and the convex function $T_3$ has maximum at one of $d,h$. The function $T_2$ is convex separately on $[d,1/2]$ and $[1/2,h]$, so it suffices to check $d,1/2,h$. The active aspect-ratio interval lies strictly above $1/2$, so only the smooth branch of $f$ enters this argument.

The exact fractions in \eqref{eq:parameters} satisfy
\[
\frac12<\frac d h<\frac1{2h}<1,\qquad
M(1/2)=h+\frac12+\frac{h^2}{2}.
\]
Consequently the remaining central endpoint is
\[
T_2(1/2)=1+\frac h2+\frac{h^2}{4}
=1.39444850340025<\tau_0=1.394449.
\]
All comparisons are exact rational comparisons; the terminating decimal for $T_2(1/2)$ is exact. The following rational upper bounds suffice at all other endpoints:
\[
\begin{array}{c|c}
\text{quantity}&\text{strict rational upper bound}\\\hline
2h&1.212\\
M(d)&1.363\\
T_3(d)&1.374\\
T_3(h)&1.357\\
T_2(d)&1.379\\
T_2(h)&1.367
\end{array}
\]
To make the endpoint comparisons fully reproducible, the six noncentral expressions simplify to
\[
\begin{aligned}
2h&=2(1-d), & M(d)&=\tfrac94(1-d),\\
T_3(d)&=\tfrac{1+d+\frac92(1-d)}3,
&T_2(d)&=\tfrac{1+d+\frac94(1-d)}2,\\
T_3(h)&=\tfrac{3+(1-d)+(1-d)^2/(2d)}3,
&T_2(h)&=\tfrac54+\frac{(1-d)^2}{8d}.
\end{aligned}
\]
All six expressions decrease strictly with $d\in(1/3,1/2)$: this is immediate for the linear terms, and for the remaining terms it follows by differentiating $(1-d)^2/d=d^{-1}-2+d$, whose derivative $1-d^{-2}$ is negative. Their displayed strict rational bounds follow by clearing positive denominators. For the limiting value $d_*=4-\sqrt{13}$ use the exact isolating inequality
\[
\frac{3944487}{10^7}<d_*<\frac{3944488}{10^7},
\]
which follows by squaring $3.6055512<\sqrt{13}<3.6055513$. Substituting the rational lower endpoint $3944487/10^7$ into the six decreasing expressions proves the same strict bounds $1.212$, $1.363$, $1.374$, $1.357$, $1.379$, and $1.367$ at $d_*$, all below $1+d_*$. The central identity $T_2(1/2)=1+h/2+h^2/4$ is exact. This supplies a rational certificate of every case, without numerical floating-point assumptions.

Each upper bound is strictly smaller than $\tau_0$. Convexity of $M$, $T_3$, and of $T_2$ separately on $[d,1/2]$ and $[1/2,h]$ establishes the uniform bound at every intermediate frequency scale. To obtain the sharp exponent from this bound, note that $d_*=4-\sqrt{13}$ and $h_*=1-d_*=\sqrt{13}-3$ satisfy
\[
1+d_*=1+\frac{h_*}{2}+\frac{h_*^2}{4}=\tau_*=5-\sqrt{13}.
\]
The strict inequalities for the other endpoints at $d_*$ follow from the exact rational isolation and monotonicity just established. Since all endpoint expressions are continuous for $1/3<d<1/2$, for each fixed $\eps>0$ we may select a rational $d_{\eps}$ sufficiently close to $d_*$ that every endpoint exponent, including the boundary cost $1+d_{\eps}$, is at most $\tau_*+\eps/2$. Choose $D=\lceil N^{d_{\eps}}\rceil$; this is exactly computable by fixed-degree integer powering and binary search as above. Give the field operations and subpolynomial factors the remaining accuracy slack. This proves the uniform running-time exponent $\tau=\tau_*$ without requiring irrational machine operations or an unproved exponent conjecture. Convexity then proves that every term in the primitive, including construction of dense operands and candidates, costs at most $N^{\tau+o(1)}$ at every intermediate scale.

There are only logarithmically many scales and product-recursion levels. Their input reads, query lists, and output writes are included. We obtain a deterministic known-threshold solver for arbitrary batches with
\begin{equation}\label{eq:threshold-time}
T_{\mathrm{thr}}(n,Q)\le(n+Q)^{\tau+o(1)}.
\end{equation}
For the uniform meaning of this statement, fix a target $\eps>0$, give the ordinary products a sufficiently small accuracy slack, and absorb the subpolynomial prime-selection overhead and logarithmic recursion factors in the remaining slack. Constants depend on $\eps$, not on input dimensions, endpoints, or colors.

\section{From threshold decisions to exact mode frequencies}\label{sec:frequencies}

For each global occurrence list of a color, retain the occurrences of even rank. The filtered array $A'$, in original order, has length at most $\lfloor n/2\rfloor$. Let $p(t)$ be the number of retained positions among the first $t$ positions of the original array. An inclusive query $[l,r]$ projects to
\[
[p(l-1)+1,p(r)],
\]
which is exactly the subsequence of retained positions lying in the query. Empty projected intervals have frequency zero.

Occurrences of any fixed color inside an interval have consecutive ranks in that color's global occurrence list. If the original count is $f$ and the retained even-rank count is $e$, then
\[
2e-1\le f\le2e+1.
\]
Let $m$ be the maximum frequency in the filtered query and $M$ the maximum frequency in the original query. A color attaining $m$ supplies the lower inequality, while applying the preceding bound to every color gives
\begin{equation}
2m-1\le M\le2m+1.
\label{eq:G8}
\end{equation}
For a nonempty original query with $m=0$, every color occurs at most once, so $M=1$. Otherwise test threshold $2m+1$. If the answer is positive, then $M=2m+1$. If it is negative, test $2m$; a positive answer gives $M=2m$, and a negative answer gives $M=2m-1$. Thus only two rounds of known-threshold batches are needed, internally divided into the dyadic scales already handled in \cref{sec:thresholds}.

The algorithm recursively solves exactly one child instance on $A'$, keeping at most $Q$ projected queries, and then resolves the original frequencies by the two threshold rounds. The array length decreases geometrically, so the recursion has $O(\log n)$ levels. Every level costs at most $N^{\tau+o(1)}$, and the logarithmic factor is absorbed into $N^{o(1)}$. Building the filter and projecting all queries costs $O(n+Q)$ per level. The empty base array has mode frequency zero. By induction, every reported frequency is exact, and the total deterministic worst-case time is
\[
N^{\tau+o(1)}.
\]

\section{Deterministic recovery of a mode color}\label{sec:decoder}

\subsection{A general decoding lemma}\label{sec:decoder-lemma}

Suppose there is a deterministic algorithm that, for every explicit array of length $m$ and every batch of $q$ intervals, returns all exact mode frequencies in time $(m+q)^{\tau+o(1)}$ for some $\tau\ge1$, including the cost of constructing and reading the instance. The frequency solver proved in \cref{sec:frequencies} has exactly this property.

\begin{lemma}[Mode-color decoding]\label{lem:decoder}
Given such an exact frequency solver, one mode color for every query on an array of length $n$ and a batch of $Q$ intervals can be recovered deterministically in worst-case time
\begin{equation}
(n+Q)^{\tau+o(1)}.
\label{eq:D}
\end{equation}
If color identifiers are totally ordered, the decoder can return the smallest identifier among all modes.
\end{lemma}

More explicitly, if the frequency solver has a uniform cost $O(M^\tau\operatorname{polylog}M)$, the decoder adds at most an $O(\log\sigma)$ factor, where $\sigma\le n$ is the number of colors. The same argument applies to subpolynomial overhead: for every $\varepsilon>0$, use a uniform bound $C_\varepsilon M^{\tau+\varepsilon}$ and absorb small instances and logarithmic factors. Original identifiers have $O(\log(n+Q))$ bits.

\subsection{Decoder algorithm}\label{sec:decoder-algorithm}

First compute the exact target frequency $M_t$ for every original query. Sort and compress the color identifiers deterministically in $O(n\log n)$ time, and build a balanced binary partition of the ordered palette. This order also fixes the canonical tie-breaking rule: always prefer the half containing smaller identifiers.

A node $v$ of the palette tree owns a subpalette $P_v$, a filtered array $A_v$ containing exactly the colors of $P_v$ in their original order, and a list of already routed queries. Each query keeps its original identifier, its target $M_t$, and its endpoints in $A_v$. Maintain the invariant that $P_v$ contains at least one mode of the original query.

Every subpalette is an interval in the root's compressed order. For local solver calls, colors may be relabeled by ranks $1,\ldots,|P_v|$ by subtracting the beginning of that interval while keeping the map to original identifiers. This costs linear time in the filtered array, already charged, and keeps local identifiers within $O(\log(|A_v|+Q_v))$ bits. The final canonical choice is still determined by the original order.

Split $P_v$ into halves $P_L$ and $P_R$. Scan $A_v$ once to build filtered arrays $A_L,A_R$ and the prefix function
\[
p(s)=\#\{j\le s:A_v[j]\in P_L\},\qquad p(0)=0.
\]
For an inclusive parent query $[l,r]$, the projected intervals are
\begin{equation}
I_L=[p(l-1)+1,p(r)],
\qquad
I_R=[l-p(l-1),\ r-p(r)].
\label{eq:P}
\end{equation}
An interval whose left endpoint exceeds its right endpoint is empty and has mode frequency zero. 
Run the deterministic frequency solver on $A_L$ for all node queries whose left projection is nonempty. If the returned frequency is $M_t$, route that query only to the left child; otherwise route it only to the right child. Equivalently, one may use the known-threshold solver to test $\modefreq(A_L,I_L)\ge M_t$. The internal solver never needs to recover a color.

When a node's palette is a singleton, that color is the answer for every query routed to the node. For nonempty original queries, $M_t\ge1$, so no routed query can end up in an empty interval. Empty original queries are handled directly according to the convention in \cref{sec:introduction}.

\subsection{Correctness of the decoder}\label{sec:decoder-correctness}

Filtering preserves occurrence order. Equation~\eqref{eq:P} selects exactly the child-color occurrences that lay in the parent interval: no outside occurrence enters and no retained occurrence inside the parent interval disappears. Hence the frequency of every retained color is unchanged by projection.

At the root, $M_t$ is the true mode frequency. Assume the routing invariant holds at a parent node. In the left child, no color can have frequency greater than $M_t$, because all left-child occurrences come from the original query. If the left solver returns $M_t$, at least one original mode lies in $P_L$, so routing left preserves the invariant. If it returns less than $M_t$, no mode lies in $P_L$; the parent invariant then forces at least one mode into $P_R$, so routing right preserves the invariant.

At a leaf, the unique remaining color therefore occurs exactly $M_t$ times in the original query and is a true mode. Always choosing the left half whenever it contains a mode discards only larger identifiers, so the output is the smallest identifier among all modes. No sampling, randomized hashing, isolation probability, or expected-time stopping rule is used.

\subsection{Worst-case cost of the decoder}\label{sec:decoder-cost}

Fix one depth of the palette tree and consider all nodes carrying active queries. Their palettes are disjoint, so their arrays are filtered from disjoint color classes of the original array. Consequently
\[
\sum_v |A_v|\le n.
\]
Every query is routed to exactly one node at that depth, hence
\[
\sum_v Q_v\le Q.
\]
The frequency solver is called on the left child of every node with at most $Q_v$ queries. Setting $s_v=|A_{L(v)}|+Q_v$ gives
\[
\sum_v s_v\le n+Q=N.
\]
For $\tau\ge1$,
\[
\sum_v s_v^\tau\le\left(\sum_v s_v\right)^\tau\le N^\tau.
\]
For a bound with $o(1)$ overhead, fix $\varepsilon>0$; then $\tau+\varepsilon\ge1$ and
\[
\sum_v C_\varepsilon s_v^{\tau+\varepsilon}
\le C_\varepsilon N^{\tau+\varepsilon}.
\]
Constructing both child arrays, the prefix vectors, and all projected queries costs $O(n+Q)$ per depth. A balanced palette partition has $O(\log\sigma)$ depths. Adding the initial frequency computation, color compression, and output writing still leaves the total within \eqref{eq:D}. Processing one depth at a time uses $O(n+Q)$ auxiliary space beyond the memory of the frequency solver, because arrays from completed depths can be discarded.

The same solver is valid on every filtered array and projected query batch because its guarantee is for arbitrary explicit arrays and arbitrary batches, with cost controlled by their total mass. At any depth, the batches separate by query and the arrays separate by palette, so summing local costs introduces no additional polynomial factor.

Combining \cref{sec:frequencies,lem:decoder} proves \cref{thm:main} for both frequencies and colors. Prime selection, data structures, recursions, and query routing are all deterministic and all stated costs are worst case. The averaging argument in \cref{sec:modulus} establishes the existence of a prime that is found by exhaustive deterministic evaluation.

\paragraph{Use of generative AI.}
Generative AI tools assisted with literature searches, drafting and editing. The author takes responsibility for all mathematical claims and conclusions

\end{document}